# Identification, Impacts, and Opportunities of Three Common Measurement Considerations when using Digital Trace Data

Daniel Muise[1,2], Nilam Ram[2,3], Thomas Robinson[4,5], & Byron Reeves[2]

Stanford University
2023

Cataloguing specific URLs, posts, and applications with digital traces is the new best practice for measuring media use and content consumption. Despite the apparent accuracy that comes with greater granularity, however, digital traces may introduce additional ambiguity and new errors into the measurement of media use. In this note, we identify three new measurement challenges when using Digital Trace Data that were recently uncovered using a new measurement framework – Screenomics – that records media use at the granularity of individual screenshots obtained every few seconds as people interact with mobile devices. We label the considerations as follows: (1) *entangling* – the common measurement error introduced by proxying exposure to content by exposure to format; (2) *flattening* – aggregating unique segments of media interaction without incorporating temporal information, most commonly intraindividually and (3) *bundling* – summation of the durations of segments of media interaction, indiscriminate with respect to variations across media segments.

**Introduction**

Almost 30 years ago, Bartels poignantly declared that media effects research was "…one of the most notable embarrassments of modern social science", in large part because of measurement error (Bartels, 1993, p. 267). Although unreliable self-reports of media exposure are still often used, we now know considerably more about the inaccuracies embedded in these measurements, especially when applied to new interactive media (Hodes & Thomas, 2021; Scharkow, 2016). Cataloguing specific URLs and applications with digital traces is the new best practice for measuring media use (Jungherr, 2018; Keusch & Kreuter, 2021). Using digital traces, researchers are measuring a wide variety of useful constructs, from total screen time (Hodes & Thomas, 2021) to diversity of engagement (Ram et al, 2020) to social media exposure (Bechmann & Nielbo, 2018; Eady, Nagler, Guess, Zilinsky, & Tucker, 2019) to news diet compositions (Guess, 2021; Muise, Hosseinmardi, Howland, Mobius, Rothschild, & Watts, 2022).

Although digital traces provide more precise measures, it would be a mistake to declare premature victory in the quest to build better records of digital life. Familiar measurement problems remain, but more importantly, the current media environment is *considerably more complex* than prior media regimes. Highly fragmented digital experiences last only seconds and are comprised of idiosyncratic threads of media use that fluidly cross traditional boundaries of content providers, platforms, and applications (Brinberg, Ram, Wang, Sundar, Cummings, Yeykelis, & Reeves, 2022; Reeves, et al., 2021, Yeykelis, Cummings, & Reeves, 2018). Despite the apparent accuracy that comes with greater granularity, digital traces may introduce additional ambiguity and new errors into the measurement of media use (Freelon, 2014; Lazer et al., 2020).

---

[1] Screenlake, Inc
[2] Department of Communication, Stanford University
[3] Department of Psychology, Stanford University
[4] Department of Pediatrics, Stanford University
[5] Department of Medicine, Stanford University

In this commentary, we identify and describe three measurement considerations which should be of interest to any communication researcher using Digital Trace Data (DTD). These considerations were recently uncovered using a new measurement framework – Screenomics – which records media-use at the granularity of individual screenshots obtained every few seconds during ordinary natural device usage (Reeves et al., 2019). This approach, described in greater detail below, offered a new level of data-richness that showed where common assumptions don't seem to hold. We will review each consideration individually by bringing Screenomics data to bear on measurement assumptions and conclusions from the contemporary DTD communication literature. We primarily illustrate these measurement issues with examples from political communication but they also apply to other substantive applications of DTD analysis. Screenomics data are uniquely suited to uncover these considerations. While not exhaustive with respect to breadth of measurement considerations within DTD analysis, these three considerations are not presently documented in the relevant literature (Cesare et al., 2018).

We label these three considerations as follows: (1) *entangling* – the common measurement error introduced by proxying exposure to content by exposure to format; (2) *flattening* – aggregating unique segments of media interaction without incorporating temporal information, most commonly intraindividually and (3) *bundling* –summation of the durations of segments of media interaction, indiscriminate with respect to variations across media segments. Each consideration is potentially relevant to a broad array of social science research beyond that which uses DTD. In this commentary, however, we frame and evaluate our findings in terms of DTD analysis due to the substantive impact these findings may have on the conclusions of contemporary large-scale DTD publications.

**Data & Method**

The data were provided by 115 study participants who were recruited via Qualtrics from around the United States to participate in a longitudinal smartphone study, as part of the Human Screenome Project (Reeves, Robinson, & Ram, 2020). Participant demographics are similar to the US population (69% white, 56% female, median age 44 years). Study participants installed screen-recording software on their smartphones for two weeks between October 2019 and January 2020. The software captured all visual screen-content every five seconds, as well as foreground-application logs. Over the course of the study, participants provided 4,907,091 individual screenshots (i.e., nearly 7000 hours of screentime) that were encrypted, compressed, and transmitted to research servers. We identified *sessions* of screen activity (defined as the time between screen on and off) by identifying gaps in the timestamp chronology of length >5s. Throughout, we use the term 'segment' more generally to refer to a continuous duration of interacting with media with a particular characteristic, be it a post, topic, video, app, etc, as identified and measured by contiguous screenshots each containing that shared characteristic.

For purposes of illustrating the impact of measurement considerations, we hereon apply our Screenomics data by commenting on the political communication literature, which is a frequent source of source of highly impactful and rigorous analysis using DTD. To do so, we identify the presence of political content throughout the screenshot record, noting that a similar procedure could be conducted with respect to other substantive domains. Using optical character recognition, all text appearing in the screenshots was extracted and placed into millions of documents, one per screenshot (Chiatti et al, 2018). Each screenshot was determined as either *containing politics* or *not containing politics*, where "politics" was defined broadly and operationalized topically. Specifically, we developed using a random forest model trained on a set of 125,473 ground-truth screenshots, hand-tagged by the lead author and seven assistants. Borrowing from Bode (2016) and Matthes et al. (2020), these ground-truth screenshots were hand-tagged as *political* if they contained textual or visual reference to any of the following topics: politicians and the U.S. presidential administration; political satire and political satirists; partisan groups; elections and campaigns; policy debates and/or decisions; political ideology and/or partisanship; hate speech, racial inequity, identity considerations or representation; terrorism, current American wars, US foreign policy, foreign affairs; analysis of or reactions to contemporary political events. Of the 125,473 tagged ground-truth screenshots, 1.99% contained political content under these parameters.

We then generated a list of 168 wide-ranging word-stems drawn from political discourse prevalent in late 2019 through a manual audit of dated news articles and public online chatter from across the political spectrum and various sources. Examples include *black.lives.matter, deport, DHS, hate.crime, hate.speech, lobbyis, locker.room.talk, president, primary.election, sanders, schiff, warren,* and *white.house,* where "." indicates a single space. Our intent was to create an oversized list not informed by our ground-truth data selection. The full list of 168 stems is provided in the Appendix. We then returned to the documents generated from each screenshot through optical character recognition (OCR), performed a simple *spellcheck* correction (Ooms, 2020), and detected the presence or absence of each of these 168 word-stems in every screenshot in our corpus. Finally, we trained and tuned a random forest model on our ground-truth data in *R*, nonparametrically using the presence or absence of each of the 168 words as features for predicting the presence or absence of *politics* in the non-tagged data. Through manifold cross-validation and iterative tuning, we achieved a satisfactory *F* score of 0.78, where *F* scores are appropriate for evaluating classification in highly skewed data.

By defining politics broadly, we were able to capture the wide range of political discourse occurring on screens while remaining conservative in testing our own *a priori* expectation that online political content consumption is rare for most individuals (Allen et al., 2020; Muise et al., 2022). By relying only on pre-defined word-stems rather than a more comprehensive and unsupervised textual approach, we avoided training our model on format-revealing features such as the name of application or artifact introduced by our OCR algorithm[6] mis-interpreting user interface features. Contiguous political screenshots with political classification comprised political content segments, with duration equal to five times the screenshot count (as we used a frame rate of 0.2 fps).

---

[6] *Pytesseract*, the Python package wrapper for Google's open-source *Tesseract* algorithm (Hoffstaetter, 2022).

## 1. Entangling, and how to disentangle

*Entangling*, the first consideration, occurs when *political* content is conflated with its traditional companion format – *news*. These are two constructs that deserve separate treatment, at least with respect to new media. In contrast to historical norms, political content can arrive from highly varied sources (e.g., friends, entertainers), and in any type of media package (e.g., Facebook posts, TikTok videos) (Ryfe, 2019). For example, there is a strong focus in political communication research on the role of social media in delivering, expediting, and corrupting political content as it is shared in social networks. It is still common practice, however, to measure the viewing and sharing of *news* articles rather than the viewing and sharing of political content that is embedded in other media packages (e.g., Muise et al., 2022).

The Screenomics data demonstrate the consequences of failing to account for the entangling consideration. Following our format-agnostic and source-agnostic classification of screenshots as *political* or not, we cross-tabulated our political classification with whether or not the foreground app at the time was a news application as defined by play store uploader categorization (e.g., the *CNN* app, the *Google News* app). For the mean participant, just 1.13% of screentime was spent on *any* news application, and just 1.68% of screentime was spent on *any* political content. And strikingly, the intersection of these two variables represented just 0.15% of total screentime, considerably less than either political words or news separately. For the average participant, only 9% of political content arrived via a news application, and more than three-fourths of time spent on news applications was spent consuming non-political content --- i.e., updates regarding sports, gossip, weather, entertainment, hyper-local affairs, and so on.

The separation of news applications and political content is not a problem of uniform bias that can be controlled statistically. The cross-tabulation for each individual study participant looks quite different. Between-person differences in the proportion of time spent on news apps and the proportion of time spent on political content was 0.37 ($df=107, p<0.001$). The heaviest consumer of news applications (in absolute terms) saw very little political content, and the participant who spent the largest share of screen time viewing political content did not receive any of it via a news app. The data highlight that the experience of words and pictures of interest in political communication cannot be accurately measured by exposure to commercial news apps, in the same way we might have assumed for offline newspapers and television news programs. The digital environment offers unlimited and easy mixing of traditional genres. Only when content and source are disentangled can theories about the effects of either category of information be tested. At best, entangled media measures represent a power problem from misclassification (i.e., measurement error) and at worst, put researchers at risk of producing incorrect results from biased measurement error. In the arms race of media technology and media measurement, new methods such as Screenomics, alongside other granular and discerning approaches, can match the richness and complexity of the flexible media platforms environments, thus allowing researchers the ability to disentangle content and format when necessary.

## 2. Flattening, and how to unflatten

The second consideration to proper media measurement, *flattening*, regards the loss of information incurred by combining segments of similarly-coded content exposure, *unweighted by duration*, into a single metric as though they were identical events. This issue is commonly found in current studies of online news article consumption (Bakshy et al, 2015; Barberá, 2014; Barberá et al., 2015; Bechmann & Nielbo, 2018; Colleoni et al., 2014; Conover et al., 2011; Eady et al, 2019; Morales et al., 2021; Munger et al., 2017; Nithyanand, Schaffner, & Gill, 2017; Peterson et al., 2021). Using logs of browser events, these studies identify, for example, an individual as partisan-left or partisan-right by tallying the number of visits to partisan-coded domains per unit of time. If a person visits ten left-leaning online publications, and five right-leaning ones, a "flattened" measure would index their use as a 2:1 left-leaning news diet. The reasonableness of this approach stems from earlier media regimes, where political content was presumed to be consumed in uniform and predictable temporal segments (e.g., 30-minute television news shows) (Lazer, 2020).

Current media, however, are consumed in highly varied temporal fragments: pausing, fast-forwarding, multiple tabs, and other interface affordances allow or entice users to consume strings of content that are highly variant in duration. In smartphone usage, segment durations commonly follow log-normal and exponential distributions, seemingly regardless of which method is used to determine the bound of a temporal segment (Brinberg, et al., 2022; Muise, et al., 2022; Rula, Jun, & Bustamente, 2015; Zhu et al., 2018). Variance in exposure duration has known implications for how information is processed, and its influence on knowledge gain and behavior (Wieland & Kleinen-von Königslöw, 2020).

Returning to our substantive example of political content measurement, the Screenomics data showed that political content segment durations indeed follow a log-normal distribution, in line with other documented segment duration distributions. In Figure 1, we show the distribution of political content segment durations for the entire sample (dark black line), and for each participant (colored lines). Almost half of all political content segments lasted just five seconds, the minimum segment duration detectable in our data (e.g., scrolling past a political Facebook post, seeing a headline or an email participant line, etc). A full 63% of observed political segments lasted for ten seconds or less. The smaller but meaningful rightward tail of the log-normal distribution stretched past five, ten, and twenty minutes in rare segments. The longest single political content segment engaged by this sample of participants (a viewing of the *Joe Rogan Experience*) lasted just under thirty minutes, 360 times longer than the modal five-second segment.

The danger of *flattening* as a measurement practice rests on the unfounded assumption that exposures are functionally equivalent. The literature suggests they are not. Different durations are linked to different processes (Lemke, 2000; Stoker, Hay, & Barr, 2016), meaning that researchers must acknowledge *some* exposures are not temporally matched to political outcomes of interest, i.e., either too short or too long to engage in a hypothesized process. Effects of longer segments may be best matched, for example, with effects that relate to rational comparison of political candidates, especially with memory for the details of issues (e.g, Ohmes, Maslowska, & Mothes, 2022; Van Erkel & Van Aelst, 2021). Effects of shorter, sub-10 second segments may be better matched with effects like intuitive impressions of politicians (Mattes et al., 2018) or the recognition of a candidate's existence based on short advertisements (Broockman & Green, 2014). At root, *flattening* fails to acknowledge the role of *time* in social science processes.

Luckily, unflattening one's media measures is fairly straightforward, provided that the data being analyzed contains temporal information, and the research being conducted provides guidance on the temporal properties of the phenomenon for which media exposure is a proxy. For example, a researcher intending to measure deliberative engagement with a given topic may want to exclude particularly brief political content segments from a count of meaningful exposure instances. We note that some otherwise rich log data sources contain only partial temporal information, such as *arrival* times at a URL but not *exit* times, subtly complicating the robust computation of durations when there may be breaks in an individual's media usage in the middle of a logging period. It is in this domain that fully comprehensive logging data such as is provided by a Screenomics framework has a clear advantage for modern media measurement.

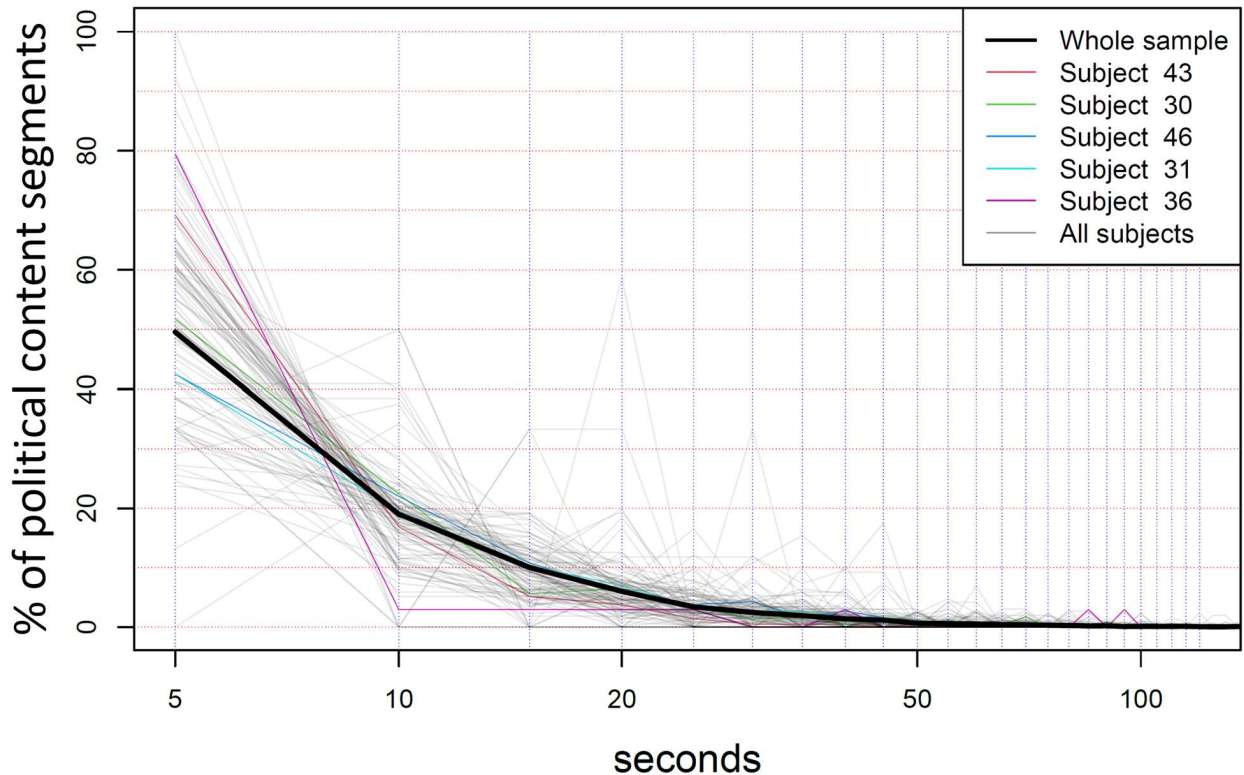

Figure 1. Distribution of durations of political content segments, for the mean of all segments pooled together (in bold black) and for each participant (in thin lines). We highlight five individual participants as well with specific colors, for illustration. On the X-axis, we show the duration of political content segments in five-second intervals. On the Y-axis, we show the percentage of political content exposure segments that lasted for that particular duration.

3. **Bundling, and how to stay un-bundled**

Whereas flattened media measures fail to incorporate temporal information, *bundled* media measures overweight temporal information by treating time as a linear commodity. Specifically, *bundling* occurs when researchers indiscrimintately sum the temporal information associated with multiple segments into a single aggregate duration metric. This commonly occurs in studies requiring a between-persons ranking of overall media exposure (e.g., Allen, Howland, Mobius, Rothschild, & Watts, 2020; Guess, 2021; Muise, Hosseinmardi, Howland, Mobius, Rothschild, & Watts, 2022). *Bundling* creates synthetic aggregates of experience that do not reflect the fragmented experiences of the individuals who experienced them. Thus, like flattening, bundled media measures assume that temporal context and durations of individual segments ultimately do not matter.

The Screenomics data easily demonstrate the potential benefits of *unbundling* aggregate durations. Suppose that for a particular measurement strategy we were to aggregate the durations of all political content segments found in our data. While this is an apparently extreme starting point, it is not dissimilar from aggregate time units summed at the monthly-level, as found in well-cited literature (e.g., Muise et al. 2022). In our aggregated unit of time, 17% is comprised of moments spent in political content segments lasting ten seconds or less. These short segments comprised more than half of all political content segments identified in our two-week study. The fastest political content segments may be uniquely pertinent to the research goals of a media user or researcher, and importantly dissimilar from longer segments. Within our data, short political content segments may represent the fast-viewing of notifications that a media consumer has delibrately subscribed to; adverse exposure to a countervailing ideology; instances of exposure embedded in contentious fast-moving feeds; or even highly-impactful repeated priming events. Each of these examples are offered to demonstrate that careful examination and interpretation of the shortest segments in isolation may be core to understanding exposure measures. In our bundled aggregate, all of these examples are reduced to just "17%". The potentially salient effects of fast experiences become relatively invisible through bundling, when in fact they may be crucial to contemporary political communication.

More broadly, bundling can be problematic for media measurement in any scenario in which segments of varied length are aggregated together (Koffer et al., 2018). The duration of a segment of media exposure is generally associated with the *intentionality* of exposure, and to the goals that media users have have for processing media (Matthes et al., 2020). To illustrate, consider a variable of high importance in political advertising: *memory of content*. Experimental research has found that a ten-second segment of exposure to advertising content begets a recognition-and-recall-rate of under 20%, versus a 97% content-recognition rate following a four minute-long exposure segment (Goldstein, McAfee, & Suri, 2011). If we apply this finding to our aggregate time-unit from the previous paragraph, we see that the inherent heterogeneity of political content segment durations severely muddles any gross interpretation of their aggregate impact on media users. If 17% of political content exposure time was in a temporal format that instills virtually no memorable impression within a media user, how can the aggregate still be treated as singular unit of content exposure?

We concede that the continuous nature of time itself demands that some level of aggregation is inevitable in any media measurement. We argue that bundling is essentially the practice of temporal aggregation done wrong. The unit we have used throughout this research note and the dissertation that inspired it, the 'political content segment', is itself a temporal aggregate composed of the initial exposure, sustained exposure, and cessation of exposure to political content, and even the term 'political content' implies aggregation across topics that our research agenda has deemed identically relevant for the purposes of our analysis. Our own aggregation decisions in this work are not infallible, but they are also far from arbitrary, and their determination is documented and justified plainly. In research making use of DTD, we suggest that the way to 'unbundle' media measures is to ensure that aggregate temporal units are considered not just in reference to the granularity of data or the ease with which overall totals can be computed, but also in reference to theoretical and conceptual goals of research. Insofar informational richness of DTD empowers rich determinations of aggregation choices, Screenomics data uniquely equips researchers for confidently unbundled analyses.

**Discussion**

Our analyses of new Screenomics data suggest the need for a look at new considerations to media measurement using DTDs. Media research has continually made advances in new environments, including more accurate logged interactions and data at scale – achievements worth celebrating. The temporal and organizational complexity of the digital world, however, and the opportunities for media users to follow idiosyncratic threads of media experience, suggest taking additional care when assessing media use.

A transition to the Screenomics data used in the present analyses provide a partial solution for DTD researchers. The richness of screen-recording data allows for precise evaluation and auditing of behavior, and for testing assumptions latent in common DTD measures. Given the difficulty of obtaining such data, it is unlikely that it can be collected at cross-sectional scales familiar to those making use of DTD, meaning that this data type is not yet a plausible alternative for many research programs. However, as open-source solutions (Yee et al., 2022) and market solutions for screen-recording become available, the incorporation of screen-recording data (or other hyper-rich data types) into larger cross-sectional DTD samples will be increasingly plausible. Hyper-rich data captured from a subset of DTD study participants or data contributors can serve as invaluable ground-truth reference for all measurement decisions.

In closing, we acknowledge the limitations of our analysis. Rich descriptive data analysis at small cross-sectional scales is necessary for highlighting potential holes in practices conducted at large scales (Molenaar & Campbell, 2009; Munger, Guess, & Hargittai, 2021), but we warn against broad generalization of *specific* findings, such as the distribution of political content across applications. Our intent is strictly to demonstrate and discuss what may have only been intuited but not yet documented by our peers, due to the paucity of sufficiently rich observational data. As inspired by excellent adjacent efforts such as Van Driel et al. (2023) and Sen et al., (2021), the applications presented in this paper are exemplary but not exhaustive. Future research may find and examine yet more applications of the considerations outlined here, or reinvestigations of past influential work with a new lens on measurement assumptions.

# Appendix

| | | | | | | |
|---|---|---|---|---|---|---|
| alex.jones | chuck.schumer | fiorina | kamala | o.rourke | quid.quo.pro | syria |
| americans | climate.change | fox.and.friends | kellyanne | obama | rachel.maddow | Taliban |
| amendment | clinton | G7 | keystone | obamacar | racis | tax |
| andrew.yang | congress | gabbard | klobuchar | ocasio.cortez | rally | terroris |
| asylum | conservativ | gavin.newsom | law.maker | ouse.majority | rand.paul | tillerson |
| ballot | constitution | george.zimmerman | lawmaker | partisan | recession | transphobic |
| barack | covfefe | gerrymand | legaliz | pelosi | reform. | trayvon.martin |
| battleground.state | debate | ginsburg | legisla | pence | representative | treason |
| beto | deep.state | giuliani | lewandowski | polariz | republican | trevor.noah |
| biden | deepstate | govern | liberal | politic | RNC | trump |
| bigly | democrat | gun.control | limbaugh | pompeo | roger.stone | tucker.carlson |
| bigot | deplorabl | hassan.minhaj | lindsey.graham | populis | sanctions | united.nation |
| black.lives.matter | deport | hate.crime | lobbyis | president | sanders | warren |
| blm | DHS | hate.speech | locker.room.talk | primary.election | schiff | white.house |
| booker | DNC | house.minority | maga. | pro.choice | senator | white.nationali |
| border.wall | dogwhistl | human.right | manafort | pro.life | shooting | white.supremac |
| breitbart | donald.trump | immigration | mandate | prochoice | shutdown | witchhunt |
| brexit | economy | impeach | medicaid | prolife | socialis | .CIA. |
| buttigieg | elected.official | impeachment | melania | prosecut | sondland | .DOJ. |
| campaign | exit.poll | incel | merkel | protest. | stephen.colbert | .election |
| capitol | extradit | incumbent | migrant | protests | supreme.court | .facis |
| castro | FBI | jeff.sessions | minorities | proud.boys | susan.collins | .GOP. |

*Table 1. A list of 168 political word stems used training our random-forest binary politics classifier. The presence or absence of each stem from a screenshot's text was used as an individual feature with which to predict whether a screenshot contains political content, based on a random forest model trained on the ground truth set. In the table, the presence of a period (.) indicates a space between characters.*

CoI notice: Author Daniel Muise is the CEO of Screenlake, Inc, a research company based on granular behavioral data collection and analysis from smartphones. The research in this note was conducted prior to and independently of Screenlake, Inc, as part of a doctoral dissertation.